# Random Linear Network Coding in NOMA Optical Wireless Networks

Ahmed A. M. Hassan, Ahmad Adnan Qidan, Mansourah K. Aljohani, Taisir E. H. El-Gorashi, Jaafar M. H. Elmirghani
School of Electronics and Electrical Engineering, University of Leeds, Leeds, United Kingdom
Email: {Elaamh, A.A.Qidan, Ml16mka, T.E.H.Elgorashi, J.M.H.Elmirghani}@leeds.ac.uk

*Abstract*— Optical wireless communication (OWC) has the potential to provide high communication speeds that support the massive use of the Internet that is expected in the near future. In OWC, optical access points (APs) are deployed on the celling to serve multiple users. In this context, efficient multiple access schemes are required to share the resources among the users and align multi-user interference. Recently, non-orthogonal multiple access (NOMA) has been studied to serve multiple users simultaneously using the same resources, while a different power level is allocated to each user. Despite the acceptable performance of NOMA, users might experience a high packet loss due to high noise, which results from the use of successive interference cancelation (SIC). In this work, random linear network coding (RLNC) is proposed to enhance the performance of NOMA in an optical wireless network where users are divided into multicast groups, and each group contains users that slightly differ in their channel gains. Moreover, a fixed power allocation (FPA) strategy is considered among these groups to avoid complexity. The performance of the proposed scheme is evaluated in terms of total packet success probability. The results show that the proposed scheme is more suitable for the network considered compared to other benchmark schemes such as traditional NOMA and orthogonal transmission schemes. Moreover, the total packet success probability is highly affected by the level of power allocated to each group in all the scenarios.

## I. INTRODUCTION

In recent years, the number of wireless devices connected to the Internet have massively increased, which causes unprecedented traffic congestion on current radio frequency (RF) wireless networks. Therefore, optical wireless communication (OWC) has received huge interest from both industrial and academic communities as a key element in the next generation of wireless communication to support the escalating demands of users for high data rates with wireless connectivity. OWC networks use optical bands at different wavelengths (colours) to send information to multiple users [1]-[4]. Thus, they provide a license-free spectrum, high energy efficiency and high security. In [5]-[8], white light emitting diodes (LEDs) are used to provide both illumination and communication. It is shown that data rates in the range of gigabits per second (Gbps) are achievable by using traditional LEDs, which are available commercially at low cost. However, as any other technology, OWC faces challenges that must be taken into consideration including the small coverage area of an optical access point (AP), which results in the need for deploying a high number of optical APs in a small area. In addition, the uplink transmission in OWC is usually carried out over frequencies other than visible light to avoid glare as the uplink sources are typically close to the user.

In general, multi-user interference in wireless communication is a crucial issue that highly dictates the achievable data rates of wireless systems. In particular, an optical wireless system must have the ability to serve multiple users simultaneously in order to maximize the spectral efficiency. In this context, various orthogonal transmission schemes such as time division multiple access (TDMA), frequency division multiple access (FDMA) and code division multiple access (CDMA) [9]–[13] were proposed to allocate exclusive resources to each user, and here any residual interference among users can be managed as noise. Despite the low complexity of these orthogonal schemes, they achieve low performance in high density optical wireless networks as the resources are divided by the total number of users. In [14], non-orthogonal multiple access (NOMA) is proposed for optical wireless networks to serve more than one user (exploiting the same resources) at a given time through controlling the power domain. Basically, in NOMA, users are classified into strong users and weak users based on their channel gains, i.e., their distance from the transmitters. Each strong user is paired with a weak user, thus forming a group. In a group, NOMA allocates a high power level to the weak user compared to the strong one in order to give each weak user the ability to decode its information directly, while the strong user preforms successive interference cancelation to decode its information. It was shown that NOMA is more suitable for optical wireless networks compared with orthogonal transmission schemes [14]. However, NOMA might suffer a high packet loss, which results from high noise between the users of the group.

At this point, random linear network coding (RLNC) is defined in [15] as a way of enhancing transmission efficiency using messages and coefficients, which can be chosen independently from a finite Galois Field (GF) [15]. In RLNC, coded packets are transmitted instead of original packets, and then users decode the original packets using for instance a Gauss-Jordan elimination process. It is shown that RLNC can efficiently improve the system capacity and reduce the number of re-transmissions in various scenarios. Moreover, RLNC does not require any code sequence synchronization and/or packet loss feedback[16]–[18]. In [19], RLNC is proposed to enhance the total packet success probability and reduce the packet delay in NOMA-based RF networks considering multicast services. In optical wireless communication, several studies have applied various network coding techniques to enhance the transmission capacity. For example, in [20] network coding is applied in the physical-layer, and in [21], [22] network coding is studied in relay-based optical wireless systems, etc. However, to the best of our knowledge, using RLNC to enhance the performance of NOMA in optical wireless communication is not addressed.

In this work, an optical wireless network is considered to serve multiple users using NOMA, while RLNC is applied to send coded packets and improve the overall performance of the network. First, we defined the system model, which contains an optical AP deployed on the ceiling to serve multiple users. The users are divided into a number of multicast groups (two groups are considered here), each multicast group is composed of users with a similar channel gain. In other words, the users located far from the AP, i.e., weak users, are more likely to belong to the same multicast group, and similarly for the strong users. Subsequently, NOMA is applied to align the interference between these multicast groups by allocating high power to the weak users belonging to a multicast group compared to the strong users belonging to the other multicast group. Finally, RLNC is applied to enhance the total success probability of the network considered. The results show that the proposed RLNC-NOMA is more suitable for optical wireless networks compared with traditional NOMA and orthogonal transmission schemes where the total success probability is enhanced considerably, which means each user can successfully decode its information with reduced errors.

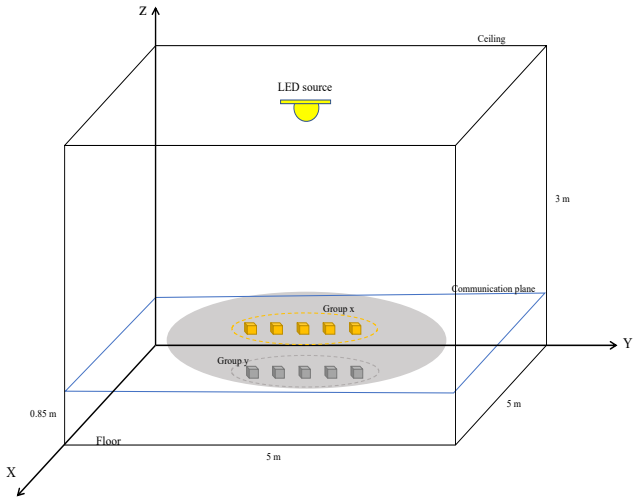

Fig. 1. OWC-system model contains two multicast groups.

The rest of this paper is organized as follows: Section II introduces the system model. In Section III, the principles of NOMA are defined. In section IV, RLNC-NOMA and RLNC-OMA are derived for the network considered. Section IV also presents the simulation setup. Section V presents the performance evaluation and results. Finally, conclusions and future work are highlighted in Section VI.

## II. SYSTEM MODEL

We consider an indoor environment as shown in Fig.1, where an optical AP is deployed on the ceiling in the center of the room to provide communication services to multiple users on the receiving plane. It is worth mentioning that the optical wireless system considered in this work is a study case to better understand the proposed schemes in the following sections. However, the models presented are general and can be further generalized to a high density optical network where a large number of APs are deployed on the celling to provide uniform coverage. The total number of users in the network is $K$. Moreover, each user is equipped with a single photo detector (PD) that points to the ceiling. Still referring to Fig. 1, users are divided into multicast groups denoted as $G$, $G=2$ in Fig.1. Each multicast group $g$ is composed of users located close to each other. In other words, in the system model, the users are classified as strong and weak users. Then, all the strong users are grouped together to form the first multicast group, while the weak users form the second multicast group.

In this work, we assume that the distribution of the users is known to the network controller, where the optical AP is connected to a central unit that controls the resources of the network and allocates them fairly among the users. In particular, NOMA is considered to align the interference between the multicast groups. Given that point, the power must be allocated between these groups in a way that gives the strong users the ability to perform SIC and decode their information, while the weak users directly decode their information managing the information intended to the strong users as noise. Therefore, we consider two levels of power allocation where a low-power level is assigned to multicast group 1 that contains the strong users (near users, ie whose channel is good), and a high-power level is allocated to multicast group 2 that contains the weak users. Note that, an optimization problem can be formulated to control power allocation among multicast groups, however it is not the aim of this work. Consequently, the power levels allocated to the multicast groups can be given as follows

$$P_1 = \alpha P_t \quad, \quad 0 < \alpha < 1 \qquad (1)$$

and

$$P_2 = P_t - P_1 \quad, \qquad (2)$$

where $\alpha$ is the NOMA power coefficient, and $P_t$ is the total power budget that guarantees uniform indoor illumination. Note that, the power allocated to multicast group 1, $P_1$, must be at least higher than zero, i.e., $P_1 > 0$, and the power allocated to multicast group 2 must be higher than $P_1$ to enforce the NOMA methodology.

The signal transmitted by the optical AP to the multicast groups can be given by

$$\mathbf{X} = x_1 + x_2 \qquad (3)$$

where $x_1$ is the information intended to the $K_1$ users that form the first multicast group, while $x_2$ contains the information intended to the $K_2$ users that form the second multicast group. Note that, each user $k$ belonging to multicast group 1 obtains packets given by the number of the users in group 1, similarly for each user $k'$, $k' \neq k$, in multicast group 2 due to the fact that multicast service is considered in this scenario. Focusing on multicast groups 1, without loss of generality, the received signal of user $k$ can be expressed as

$$y_k = h_k x_1 + z_k \qquad (4)$$

where $h_k$ is the channel of user $k$, and $z_k$ represents real valued additive white Gaussian noise (AWGN) with zero mean and variance $\sigma^2 = N_0 B$, where $N_0$ is the noise power spectral density and $B$ represents the channel bandwidth. It is worth mentioning that the received signal of user $k$ is expressed after preforming SIC and removing the information intended to the users of multicast group 2, and therefore, its information can be decoded by solving equation (4). In the following, the transmitter, channel model and receiver are introduced in detail.

## A. Optical transmitter

There are many types of light sources that can be used as transmitters. In this work, LED is used to send information to the users as shown in Fig. 1. In OWC, the data is modulated over the intensity of the radiated light such that the illumination function of the LED is not affected, and uniform illumination is achieved. Furthermore, Intensity Modulation/Direct Detection (IM/DD) is considered in our OWC system, which means that the transmitted signal must be real-valued and non-negative. Therefore, a DC bias current must be applied to eliminate the negative values of the transmitted signal. It is worth pointing out that transmitting the signal without ensuring its non-negativity might cause clipping distortion to useful information, which then results in data rate degradation in the network.

Transmitting information out of the linear range of the LED (ie large signal current variations, as the LED transfer characteristics are non-linear in any case) generates high distortion on the symbols transmitted to users regardless of the modulation technique. Therefore, we assume a limited drive current range $[I_L; I_H]$ that allows the LED operation to be modelled as linear, where $I_L$ is the turn on current of the LED, while $I_H$ corresponds to the maximum input that ensures both the linear response of the LED and the human eye safety. Within this range, the LED provides a range of output powers given by $[P_{min}; P_{max}]$. In our system model, the total power budget of the LED is $P_t$ (see equation (2)).

## B. Optical channel model

The optical channel in an indoor environment is given by line-of-sight (LoS) and diffuse components. The LoS received power usually represents 90% of the overall received power since it is a direct link between the user and the AP. The diffuse components are received due to reflections from the walls of the room and any other objects. Note that, we assume that each user is equipped with an optical detector that points towards the optical AP. Therefore, the diffuse components are neglected for the sake of simplicity. Assuming that the distance between the optical AP and user $k$ is denoted as $d_k$, the LoS component between the optical AP and user $k$ at its photodiode can be expressed as

$h_k =$

$$\begin{cases} \dfrac{(m+1)A_{PD}}{2\pi d_k^2} \cos^m(\emptyset_k)\cos(\theta_k), & 0 \leq \theta_k \leq \psi_c \\ 0, & \theta_k > \psi_c \end{cases} \quad (5)$$

where $\emptyset_k$ and $\theta_k$ donate the irradiance and incidence angles between the optical AP and user $k$, respectively, $\psi_c$ represents the field of view (FoV) angle of the receiver. Moreover, $A_{PD}$ denotes the effective area of the photodiode and $m = -1/log_2\left(\cos\left(\varphi_{\frac{1}{2}}\right)\right)$ is the LED Lambert index with a semi-angle at half-power denoted by $\varphi_{\frac{1}{2}}$.

## III. NON-ORTHOGONAL MULTIPLE ACCESS (NOMA)

NOMA is a promising multiple access technique proposed first for RF networks to allow serving multiple users using the same resources. In [14], NOMA is applied also in optical wireless networks where high performance is achieved compared to orthogonal transmission schemes such as TDMA and FDMA. In contrast, NOMA is used in this work to avoid the interference among multicast groups. In the following, the principles of NOMA are introduced for multi-users scenarios in order to model the proposed scheme in Section IV.

Considering the implementation of NOMA in optical wireless networks, users are divided into multiple groups $G$, each group contains two users that differ in their channel gains, i.e., each group contains weak and strong users. Note that, an algorithm referred to as next-largest-difference user pairing (NLUPA) can be implemented for this purpose where users are sorted in ascending order from lower to higher channel gains as $h_1 \ldots \leq h_k \ldots \leq h_K$[14]. It is worth pointing out that the interference among the groups is usually avoided by implementing an orthogonal transmission scheme, which divides the bandwidth by the number of the groups, allocating an exclusive frequency slot to each group. Focussing on group $g$, the transmitted signal intended to that group is given by

$$X_g = \sum_{k=1}^{K_g} \sqrt{P_k}\, x_k + I_{dc} \quad (6)$$

where $x_k$ is the transmitted signal to user $k$, $P_k$ is the power allocated to user $k$, and $K_g$ is the total number of the users belonging to group $g$. Moreover, $I_{dc}$ is the direct bias current applied to ensure the non-negativity of the transmitted signal. Note that, the transmitted signal must be subject to the following constraints

$$\sum_{k=1}^{K} \sqrt{P_k} \leq I_{dc} \quad (7)$$

and

$$\sum_{k=1}^{K} \sqrt{P_k} \leq A - I_{dc} \quad (8)$$

where A is the maximum optical intensity, to satisfy the non-negativity condition and the eye safety constraint. On the user's side, each strong user performs SIC to decode its information, which usually has a low power level compared to the information intended to weak users. Deeply speaking, each strong user decodes first the information of other users with high power levels following a decoding order starting from the signal with the highest power until the desired signal is decoded. Note that, the accuracy of SIC is determined by several factors including power allocation and the quality of the channel state information (CSI). On the other hand, each weak user can decode its information directly managing the information of other users with low power levels as noise. At this point, the received signal of user $k$ belonging to group $g$ is given by

$$y_k = h_k \sqrt{P_k}\, x_k + z_k \quad (9)$$

where $z_k$ is the overall noise that includes the noise that results from subtracting the information of other users based on NOMA. Considering that user $k$ belonging to group $g$ is a weak user, its achievable data rate after solving equation (9) is given by

$$r_k = \frac{B}{G}\log_2(1 + \frac{\mathcal{R}_k^2 h_k^2 P_k}{\sum_{k'}^{K_g} \mathcal{R}_k^2 h_k^2 P_{k'} + \sigma^2}) \quad (10)$$

where $k \neq k'$. Note that, the bandwidth is divided by the number of groups to avoid inter-group interference. The use of NOMA in optical wireless communication is straightforward, and it has low complexity compared to other advanced multiple access schemes. However, using NOMA to serve multiple multicast groups is insufficient if the noise due to applying SIC is high, which may lead to severe degradation in the performance in terms of packet loss. Therefore, in the next section, Random Linear Network Coding is proposed to generate coded packets following a certain method in order to maintain the original packets of NOMA without distortion.

IV. RANDOM LINEAR NETWORK CODING-BASED NOMA

In this section, RLNC is applied to enhance the success probability of NOMA-based transmission. In particular, RLNC uses encoding coefficients to encode the original packets intended for transmission using NOMA to generate coded packets to multiple multicast groups. It is worth mentioning that these encoding coefficients are unique for each multicast group, and they are linearly independent among them. Moreover, the decoding probability is 1 due to the fact that the encoding coefficients might be generated in the Galois Field GF ($2^8$), more details and mathematical expressions are derived in [16]-[17]. Interestingly, the transmission opportunity in RLNC-based NOMA transmission must be almost the same as that of NOMA transmission. Therefore, it is assumed that the total number of coded packets is determined taking into consideration the number of NOMA packets in a frame. Appling the methodology above on our system model, the achievable data rate of each user $k$ belonging to multicast group 1 containing the strong users is given by

$$R_{1,k} = B\log_2\left(1 + \frac{\mu_k \mathcal{R}_k^2 h_k^2 P_1}{\sigma_k^2}\right) \leq \delta \quad (11)$$

While the achievable user rate of each user $k'$ belonging to multicast group 2 containing the strong users is given by

$$R_{2,k'} = B\log_2\left(1 + \frac{\mu_{k'} \mathcal{R}_{k'}^2 h_{k'}^2 P_2}{\mathcal{R}_{k'}^2 h_{k'}^2 P_1 + \sigma_{k'}^2}\right) \leq \delta \quad (12)$$

where $\delta$ is the capture ratio of the packets, $\mathcal{R}$ is the photodetector responsivity of each user in A/W, $\mu$ is a constant to maintain the proportionality. It is worth pointing out that NOMA in equations (11) and (12) is used to avoid the interference between the two multicast groups.

The process of RLNC-based NOMA transmission is achieved through three steps: 1) choosing linearly independent encoding coefficients, 2) generating coded packets, 3) allocating different power levels according to equation (1) and (2) to the superposed packets, 4) users decode their messages using SIC for NOMA and a Gauss-Jordan elimination process for RLNC. In the following, the total success probability for the proposed scheme, RLNC-based NOMA, is derived.

A. RLNC-NOMA success probability formulation

In this sub-section, the total success probability of RLNC-based NOMA transmission is derived to evaluate the performance of the system model considered in section II. Basically, the total success probability is defined as the proportion of users that recover all packets correctly in a limited number of transmissions. First, the number of successive transmissions is given by $V$. Then, to derive the total success probability, the failure probability to transmit an arbitrary packet to a generic user $k$ can be expressed as follows

$$\delta_{k,v} = 1 - \exp(-\varepsilon_k) \sum_{v=1}^{V} \frac{(\varepsilon_k)^{v-1}}{(v-1)!}, \forall k \in \{K_1, K_2\} \quad (13)$$

where, $v$ is the transmission index at which the successive transmission occurred. Moreover, $\varepsilon$ denotes the ratio of the optical channel gain to the power allocated to user $k$, which can be derived for each user in multicast group 1 and multicast group 2, as follows

$$\varepsilon_{1,k} = \frac{G_{1,k}}{P_1} \quad (14)$$

and

$$\varepsilon_{2,k'} = \frac{G_{2,k'}}{P_2 - P_1(2^{(\delta/B)} - 1)} \quad (15)$$

respectively, where $G_{1,k}$ and $G_{2,k'}$ are the channel gains of the strong and weak users, which can be derived easily from equations (11) and (12).

To compute the success probability of each user belonging to either multicast group 1 or 2, first it is assumed that the optical source transmits each packet twice in the NOMA case and four coded packets in the case of RLNC-based NOMA in order to maintain fairness. Second, the binomial probability density function should satisfy the minimum reception of each user, which is at least two packets in both scenarios. Accordingly, the total success probabilities for both cases at $v$th successive transmission are expressed as

$$P_S^{NOMA} = \prod_{k=1}^{K_1}(1-\delta_{k,v})^f \prod_{k'=1}^{K_2}(1-\delta_{k',v})^f \quad (16)$$

$$P_S^{RLNC}_{NOMA} = \prod_{k=1}^{K_1}\left[1 - \left(\sum_{i=0}^{f-1}\binom{\tau}{i}(\delta_{k,1})^{\tau-i}(1-\delta_{k,1})^i\right)\right]$$

$$\times \prod_{k'=1}^{K_2}\left[1 - \left(\sum_{i=0}^{f-1}\binom{\tau}{i}(\delta_{k',1})^{\tau-i}(1-\delta_{k',1})^i\right)\right]$$

(17)

where $f$ is the number of packets in the frame and $\tau$ donates the total number of transmission attempts to multicast groups 1 and 2 in the case of RLNC-based NOMA. It is worth pointing out that $v = 1$ in equation (20) due to the fact that users receive a linearly independent coded packet at given time.

*B. RLNC-OMA success probability formulation*

For baseline analysis, we consider the use of RLNC in OMA-based transmission where the bandwidth is dived by the number of multicast groups to avoid the interference among groups. Focussing on our system model in Section II, the achievable user rates in multicast groups 1 and 2 are given by

$$R^O_{1,k} = \frac{B}{G} \log_2\left(1 + \frac{\mu_k \mathcal{R}^2_k h^2_k P_1}{\sigma^2_k}\right) \leq \delta \quad (18)$$

and

$$R^O_{2,k'} = \frac{B}{G} \log_2\left(1 + \frac{\mu_{k'} \mathcal{R}^2_{k'} h^2_{k'} P_2}{\sigma^2_{k'}}\right) \leq \delta \quad (19)$$

respectively, where $G = 2$, hence the bandwidth is divided equally devoting an exclusive frequency slot to each multicast group. Therefore, in equations (18) and (19) the interference between the multicast groups is zero, and the achievable user rate is only affected by the level of noise in the network. At this point, the total packet failure probability for a generic user $k$, until the $vth$ successive transmission occurred, is given by

$$\delta^o_{k,v} = 1 - \exp(-\varepsilon^o_k) \sum_{v=1}^{V} \frac{(\varepsilon^o_k)^{v-1}}{(v-1)} \quad (20)$$

where $\varepsilon^o_{1,k} = \frac{(B/2)G_{1,k}}{P_1}$ for each user $k$ belonging to multicast group 1, and $\varepsilon^o_{2,k'} = \frac{(B/2)G_{2,k'}}{P_2}$ for each user $k'$ belonging to multicast group 2. Following the same mathematical fashion as in the previous sub-section, the total packet success probabilities for the case of RLNC-based OMA transmission can be easily derived.

TABLE I.   SYSTEM PARAMETERS

| Parameters | Configurations |
|---|---|
| Room size | 5 x 5 x 3 m³ |
| Cell size | 3.6 m |
| LED power | 1 Watt |
| Half-power angle | 60° |
| LED Transmitter location | (2.5,2.5,3) |
| Users equipment height | 0.85 m |
| Reciever Field of View angle (FoV) | 35° |
| Responsivity of PD (A/W) | 0.4 A/W |
| Number of users in each multicast group | 5 users |
| Near group locations (x,y,z) | (1.5,2.5,0.85), (2,2.5,0.85), (2.5,2.5,0.85), (3,2.5,0.85), (3.5,2.5,0.85), |
| Far group locations (x,y,z) | (2.5,3,0.85), (3,3,0.85), (3.5,3,0.85), (4,3,0.85), (4.5,3,0.85) |
| Elevation | 90° |
| Azimuth | 0° |
| Active area of PD | 1 cm2 |
| Modulation bandwidth | 20 MHz |
| Noise power spectral density | 10-21 W/Hz |
| Frame size | 3 packets |
| Minimum NOMA user througphut | 0.5 bits/s.Hz-1 |

## V. RESULTS AND DISCUSSION

In this section, we evaluate the performance of the proposed schemes in terms of total packet success probability in an OWC indoor environment. Our analysis focuses on the relationship between packet success probability and the NOMA power coefficient $\alpha$, which determines the power ratio allocated to each multicast group, since it is the vital parameter in NOMA. Note that, the strong users in our system model decode the information intended to both groups using SIC in the case of NOMA and SIC and Gaussian elimination process in the RLNC-based NOMA case. As in the system model, two groups are considered, while each group is composed of 5 users. Moreover, the frame size is 3, and all the other parameters are listed in Table 1.

In Fig.2, the total success probability of the proposed RLNC-NOMA is depicted versus the power coefficient $\alpha$. It can be seen that the RLNC-NOMA scheme provides high success probability compared to traditional NOMA and OMA schemes, which can be interpreted as that the proposed scheme is more resistant in high noise scenarios, which is expected due to the fact that the original packets are superposed into coded packets that reduce packet distortion. Note that, the highest success probability for RLNC-NOMA is given at $\alpha = 0.33$, which matches the optimal power allocation in NOMA in [23].

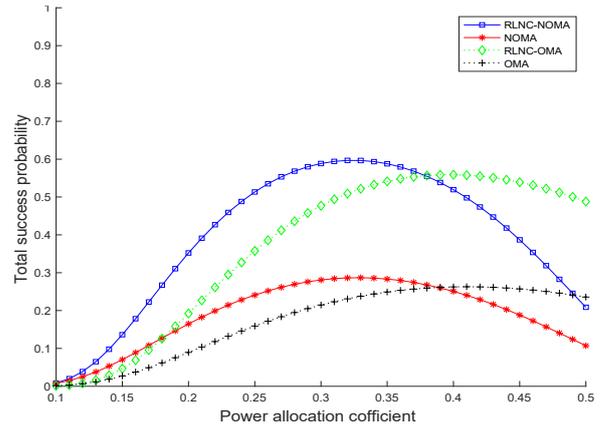

Fig. 2.   RLNC-NOMA and RLNC-OMA packet success probability versus the power allocation coffcient in OWC.

The figure further shows that RLNC-NOMA provides a total packet success probability that decreases as the power coefficient increases beyond 0.4 due to high interference between the multicast groups. As a comparison between the total success probabilities of the proposed RLNC-NOMA and RLNC-OMA schemes, it can be easily seen that both RLNC-NOMA and RLNC-OMA provide a higher success probability compared to traditional NOMA and OMA schemes. Moreover, RLNC-NOMA achieves higher success probability than RLNC-OMA for a power coefficient range from 0.1 to 0.35. However, RLNC-OMA outperforms RLNC-NOMA in high interference scenarios, where in RLNC-OMA, an exclusive frequency slot is assigned to each multicast group avoiding the interference among them. However, RLNC-NOMA is worth implementing since it provides high achievable data rates compared to RLNC-OMA especially when the power is allocated in an optimal fashion among the groups.

Fig.3 shows the performance of NOMA and OMA schemes in terms of the sum rate of the network. Recall that the power allocated to each group in our system model is determined by the power coefficients $\alpha$ (see equations (1) and (2)). Therefore, the sum rate of each group changes considerably as the power coefficient changes. In the NOMA scenario, it is shown the sum rate of multicast group 1 increases with the power coefficient, while the sum rate of multicast group 2 decreases due to severe interference between the groups. Moreover, NOMA provides high sum rates in all the scenarios considered compared to OMA, in which the bandwidth is divided by the number of the groups in the network.

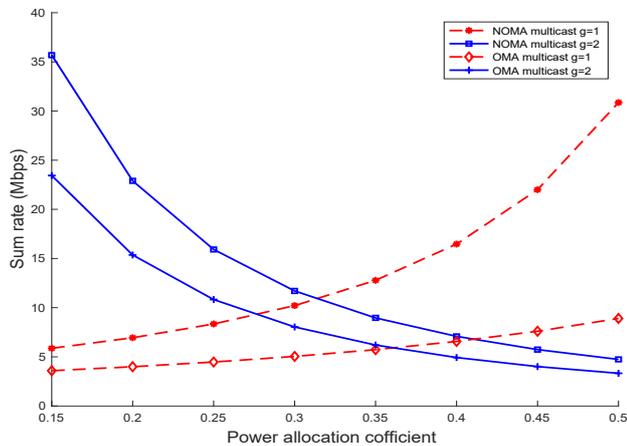

Fig. 3. NOMA and OMA sum rates versus the power allocation cofficient for two multicast groups in OWC.

## VI. Conclusions and future work

In this work, RLNC is applied to enhance the performance of NOMA in optical wireless communication. We first define a multicast system model that contains multiple users that differ in their channel gains. The users are divided into two multicast groups where each group contains users that slightly differ in their channel gains. Subsequently, the total success probability equations are derived for RLNC-NOMA and RLNC-OMA. The results show that the proposed schemes provide high success probability compared to traditional NOMA and OMA schemes. As Future work, power allocation optimization will be considered to maximize the RLNC decoding performance in realistic optical wireless configurations.

## Acknowledgment

This work has been supported by the Engineering and Physical Sciences Research Council (EPSRC), in part by the INTERNET project under Grant EP/H040536/1, and in part by the STAR project under Grant EP/K016873/1 and in part by the TOWS project under Grant EP/S016570/1. All data are provided in full in the results section of this paper.

## References

[1] O. Alsulami, A. T. Hussein, M. T. Alresheedi, and J. M. H. Elmirghani, ''Optical Wireless Communication Systems, A Survey'', p. 22.

[2] P. H. Pathak, X. Feng, P. Hu, and P. Mohapatra, ''Visible Light Communication, Networking, and Sensing: A Survey, Potential and Challenges'', IEEE Commun. Surv. Tutorials, vol. 17, no. 4, Art. no. 4, 2015, doi: 10.1109/COMST.2015.2476474.

[3] L. E. M. Matheus, A. B. Vieira, L. F. M. Vieira, M. A. M. Vieira, and O. Gnawali, ''Visible Light Communication: Concepts, Applications and Challenges'', IEEE Commun. Surv. Tutorials, vol. 21, no. 4, Art. no. 4, 2019, doi: 10.1109/COMST.2019.2913348.

[4] J. R. Barry, J. M. Kahn, W. J. Krause, E. A. Lee, and D. G. Messerschmitt, ''Simulation of multipath impulse response for indoor wireless optical channels'', IEEE J. Select. Areas Commun., vol. 11, no. 3, Art. no. 3, Apr. 1993, doi: 10.1109/49.219552.

[5] J. M. Kahn and J. R. Barry, ''Wireless Infrared Communications'', WIRELESS INFRARED COMMUNICATIONS, vol. 85, no. 2, Art. no. 2, 1997.

[6] A. G. Al-Ghamdi and J. M. H. Elmirghani, ''Spot diffusing technique and angle diversity performance for high speed indoor diffuse infra-red wireless transmission'', IEE Proceedings - Optoelectronics, vol. 151, no. 1, pp. 46–52, Feb. 2004, doi: 10.1049/ip-opt:20040140.

[7] F. Alsaadi and J. Elmirghani, ''Mobile Multigigabit Indoor Optical Wireless Systems Employing Multibeam Power Adaptation and Imaging Diversity Receivers'', Journal of Optical Communications and Networking, vol. 3, no. 1, pp. 27–39, Jan. 2011, doi: 10.1364/JOCN.3.000027.

[8] A. T. Hussein and J. M. H. Elmirghani, ''10 Gbps mobile visible light communication system employing angle diversity, imaging receivers, and relay nodes'', IEEE/OSA Journal of Optical Communications and Networking, vol. 7, no. 8, pp. 718–735, Aug. 2015, doi: 10.1364/JOCN.7.000718.

[9] S. M. R. Islam, N. Avazov, O. A. Dobre, and K. Kwak, ''Power-Domain Non-Orthogonal Multiple Access (NOMA) in 5G Systems: Potentials and Challenges'', IEEE Commun. Surv. Tutorials, vol. 19, no. 2, pp. 721–742, 2017, doi: 10.1109/COMST.2016.2621116.

[10] S.-M. Kim, M.-W. Baek, and S. H. Nahm, ''Visible light communication using TDMA optical beamforming'', J Wireless Com Network, vol. 2017, no. 1, Art. no. 1, Dec. 2017, doi: 10.1186/s13638-017-0841-3.

[11] H. Marshoud, P. C. Sofotasios, S. Muhaidat, G. K. Karagiannidis, and B. S. Sharif, ''On the Performance of Visible Light Communication Systems With Non-Orthogonal Multiple Access'', IEEE Transactions on Wireless Communications, vol. 16, no. 10, pp. 6350–6364, Oct. 2017, doi: 10.1109/TWC.2017.2722441.

[12] Y. Cai, Z. Qin, F. Cui, G. Y. Li, and J. A. McCann, ''Modulation and Multiple Access for 5G Networks'', IEEE Commun. Surv. Tutorials, vol. 20, no. 1, Art. no. 1, 2018, doi: 10.1109/COMST.2017.2766698.

[13] S. Al-Ahmadi, O. Maraqa, M. Uysal, and S. M. Sait, ''Multi-User Visible Light Communications: State-of-the-Art and Future Directions'', IEEE Access, vol. 6, pp. 70555–70571, 2018, doi: 10.1109/ACCESS.2018.2879885.

[14] S. Feng, R. Zhang, W. Xu, and L. Hanzo, ''Multiple Access Design for Ultra-Dense VLC Networks: Orthogonal vs Non-Orthogonal'', IEEE Trans. Commun., vol. 67, no. 3, Art. no. 3, Mar. 2019, doi: 10.1109/TCOMM.2018.2884482.

[15] R. Ahlswede, N. Cai, S.-Y. R. Li, and R. W. Yeung, ''Network information flow'', IEEE Transactions on Information Theory, vol. 46, no. 4, pp. 1204–1216, Jul. 2000, doi: 10.1109/18.850663.

[16] T. Ho et al., ''A Random Linear Network Coding Approach to Multicast'', IEEE Transactions on Information Theory, vol. 52, no. 10, pp. 4413–4430, Oct. 2006, doi: 10.1109/TIT.2006.881746.

[17] C.-F. Chiasserini and C. Casetti, ''Decoding Probability in Random Linear Network Coding with Packet Losses'', IEEE COMMUNICATIONS LETTERS, vol. 17, no. 11, p. 4, 2013.

[18] H.-T. Lin, Y.-Y. Lin, W.-S. Yan, and F.-B. Hsieh, ''Efficient Frame Aggregation Transmission Using Random Linear Network Coding'', IEEE Wireless Communications Letters, vol. 3, no. 6, pp. 629–632, Dec. 2014, doi: 10.1109/LWC.2014.2351408.

[19] S. Park and D.-H. Cho, ''Random Linear Network Coding Based on Non-Orthogonal Multiple Access in Wireless Networks'', IEEE Communications Letters, vol. 19, no. 7, pp. 1273–1276, Jul. 2015, doi: 10.1109/LCOMM.2015.2436392.

[20] Y. Hong, L.-K. Chen, and X. Guan, ''Adaptive physical-layer network coding over visible light communications'', in 2017 Optical Fiber Communications Conference and Exhibition (OFC), Mar. 2017, pp. 1–3.

[21] Z.-Y. Wang, H.-Y. Yu, and D.-M. Wang, ''Energy-Efficient Network Coding Scheme for Two-Way Relay Visible Light Communications'', in 2018 IEEE 18th International Conference on Communication Technology (ICCT), Oct. 2018, pp. 310–315. doi: 10.1109/ICCT.2018.8600031.

[22] Y. Hong, L.-K. Chen, and J. Zhao, ''Channel-Aware Adaptive Physical-Layer Network Coding Over Relay-Assisted OFDM-VLC Networks'', Journal of Lightwave Technology, vol. 38, no. 6, pp. 1168–1177, Mar. 2020, doi: 10.1109/JLT.2019.2954401.

[23] H. Marshoud, V. M. Kapinas, G. K. Karagiannidis, and S. Muhaidat, ''Non-Orthogonal Multiple Access for Visible Light Communications'', IEEE Photonics Technology Letters, vol. 28, no. 1, pp. 51–54, Jan. 2016, doi: 10.1109/LPT.2015.2479600.